\documentclass[12pt]{article}
\usepackage{epsfig}
\catcode`@=11

\def\@citex[#1]#2{\if@filesw\immediate\write\@auxout{\string\citation{#2}}\fi
  \def\@citea{}\@cite{\@for\@citeb:=#2\do
    {\@citea\def\@citea{,\penalty\@m}\@ifundefined
      {b@\@citeb}{{\bf ?}\@warning
       {Citation `\@citeb' on page \thepage \space undefined}}%
\hbox{\csname b@\@citeb\endcsname}}}{#1}}

\def\citer{\@ifnextchar [{\@tempswatrue\@citexr}{\@tempswafalse\@citexr[]}}

\def\@citexr[#1]#2{\if@filesw\immediate\write\@auxout{\string\citation{#2}}\fi
  \def\@citea{}\@cite{\@for\@citeb:=#2\do
    {\@citea\def\@citea{--\penalty\@m}\@ifundefined
       {b@\@citeb}{{\bf ?}\@warning
       {Citation `\@citeb' on page \thepage \space undefined}}%
\hbox{\csname b@\@citeb\endcsname}}}{#1}}

\begin{document}
 
\thispagestyle{empty}
\begin{flushright}
MPI--PHT--2002--19\\
LMU 02/04\\
August 2002
\end{flushright}

\vspace*{1.5cm}
\centerline{\Large\bf The Double Radiative Decays $B \to \gamma\gamma$}
\vspace*{0.3cm}
\centerline{\Large\bf in the Heavy Quark Limit }
\vspace*{2cm}
\centerline{{\sc Stefan W. Bosch}\ ${}^{a}$ and 
{\sc Gerhard Buchalla}\ ${}^b$}
\bigskip
\bigskip
\centerline{\sl ${}^a$ Max-Planck-Institut f\"ur Physik,
 Werner-Heisenberg-Institut,}
\smallskip
\centerline{\sl F\"ohringer Ring 6, D-80805 Munich, Germany}
\medskip
\centerline{\sl ${}^b$ Ludwig-Maximilians-Universit\"at M\"unchen, 
Sektion Physik,} 
\smallskip
\centerline{\sl Theresienstra\ss e 37, D-80333 Munich, Germany}

\vspace*{2.5cm}
\centerline{\bf Abstract}
\vspace*{0.3cm}
We analyze the double radiative $B$-meson decays $B_s\to \gamma\gamma$ and 
$B_d\to\gamma\gamma$ in QCD factorization based on the heavy-quark limit 
$m_b\gg \Lambda_{QCD}$. We systematically discuss the various contributions 
to these exclusive processes. 
The dominant effect arises from the magnetic-moment type
transition $b\to s(d)\gamma$ where an additional photon is emitted
from the light quark (one-particle reducible diagram).
The contributions from one-particle irreducible diagrams are 
power suppressed. We argue that they are still calculable within QCD 
factorization. They are used to compute the CP-asymmetry in 
$B\to\gamma\gamma$ and to estimate so-called 
long-distance contributions in $B$ and $D\to\gamma\gamma$. 
Numerical results are presented for branching ratios and
CP asymmetries.

\noindent 
\vfill

\newpage
\pagenumbering{arabic}

\section{Introduction}\label{sec:intro}
The rare decays of $B$ mesons provide an excellent means to further explore 
the Standard Model (SM) and to detect New Physics. 
In particular the Cabibbo-favoured radiative $b\to s\gamma$ modes belong 
to the small number of rare decays that are experimentally accessible already 
at present \citer{CHEN,AUB1} and give severe bounds on the parameter space 
of New Physics scenarios. 
On the theoretical side both the inclusive \citer{AY,CMM} and exclusive 
\citer{BB,AP} $b\to s\gamma$ decays are now known at 
next-to-leading-logarithmic (NLL) accuracy.

For the double radiative decays $B\to\gamma\gamma$ experimentally so far 
only upper limits on the branching fractions exist:
\begin{equation}\label{brbsdggex}
\begin{array}{rclll}
B(B_s\to\gamma\gamma) & < & 1.48\cdot 10^{-4} & \mbox{at 90\% C.L.} & 
\cite{L3}\\
B(B_d\to\gamma\gamma) & < & 1.7\cdot 10^{-6} & \mbox{at 90\%  C.L.} & 
\cite{AUB2}
\end{array}
\end{equation}
The Standard Model expectations are roughly two orders below the current 
upper limits. 
These decays have a rather clean experimental signature.
Furthermore they are of interest because they could provide useful tests
of QCD dynamics in $B$ decays.
The $B\to\gamma\gamma$ modes realize the exceptional situation of
nontrivial QCD dynamics related to the decaying $B$, in conjunction with
a completely nonhadronic final state and simple two-body kinematics.
In principle they also probe the CKM parameters $V_{ts}$ and $V_{td}$
and could allow us to study CP violating effects as the two-photon system
can be in a CP-even or CP-odd state. While contributions
of New Physics to $B_s\to\gamma\gamma$ are strongly constrained
by the measurements of $b\to s\gamma$ transitions, a sizable
enhancement of $B_d\to\gamma\gamma$ from effects beyond the Standard
Model is still conceivable.

The theoretical treatments of $B_s\to\gamma\gamma$ performed so far 
\citer{LLY,AT} all had to employ hadronic models to describe the $B_s$ 
meson bound state dynamics. A clear separation of short- and long-distance 
dynamics and a distinction of model-dependent and model-independent features
were therefore not possible. This concerns especially the dynamics of the
light-quark constituent inside the $B$, but also contributions from 
intermediate $J/\psi$, $\eta_c$, $\phi$  or $D_s^{(*)}$ meson states, 
which have been discussed as sources of potentially important
long-distance effects.
Another process related to $B\to\gamma\gamma$ is the decay
$B\to l^+l^-\gamma$. This decay has recently been studied in \cite{DS}.

We present in this paper a systematic analysis of the exclusive double 
radiative decays $B_{s,d}\to\gamma\gamma$ in QCD, based on the heavy quark 
limit $m_b\gg\Lambda_{QCD}$. This limit allows us to give a factorization 
formula for the hadronic matrix elements of local operators in the 
weak Hamiltonian \cite{BB,BFS,AP,BBNS}. In this manner we can systematically 
separate perturbatively calculable hard scattering kernels from the 
nonperturbative $B$-meson light-cone distribution amplitude (LCDA). 
Power counting in $\Lambda_{QCD}/m_b$ allows us to identify leading and 
subleading contributions to $B\to\gamma\gamma$. Only one diagram contributes 
at leading power, but the most important subleading contributions can also 
be calculated. The inclusion of these corrections is used to estimate 
CP asymmetries in $B\to\gamma\gamma$ and 
one-particle irreducible two-photon emission from light-quark loops
in $D\to\gamma\gamma$. These corrections represent the quark-level
analogue of so-called long-distance contributions in these decays.

The remainder of this paper is organized as follows. 
In section \ref{sec:basics} we give the basic formulas necessary for the 
calculation of the observables of interest. Section \ref{sec:ldgpow} contains
our discussion of leading-power and section \ref{sec:subldgpow} the one of 
subleading-power contributions.
In section \ref{sec:ldcont} we comment on long-distance contributions in
$B\to\gamma\gamma$ and $D\to\gamma\gamma$ decays.
Numerical results are presented and discussed in section \ref{sec:phen}. 
Section \ref{sec:concl} contains our conclusions.

\section{Basic formulas}\label{sec:basics}
The effective Hamiltonian for $b\to s\gamma\gamma$ is identical to the one 
for $b\to s\gamma$ transitions. This is because the equations of motion (EOM) 
can be used to reduce the basis with operators containing two photon fields 
\cite{GSW}. Up to corrections of order $1/M_W^2$ the effective Hamiltonian 
thus reads
\begin{equation}\label{heff}
{\cal H}_{eff}=\frac{G_F}{\sqrt{2}}\sum_{p=u,c}\lambda_p^{(s)}
\left[ C_1 Q^p_1 + C_2 Q^p_2 +\sum_{i=3,\ldots ,8} C_i Q_i\right]+h.c.
\end{equation}
where
\begin{equation}\label{lamps}
\lambda_p^{(s)}=V^*_{ps}V_{pb}
\end{equation}
The operators are given by
\begin{eqnarray}\label{q1def}
Q^p_1 &=& (\bar sp)_{V-A}(\bar pb)_{V-A} \\
Q^p_2 &=& (\bar s_i p_j)_{V-A}(\bar p_j b_i)_{V-A} \\
Q_3 &=& (\bar sb)_{V-A} \sum_q (\bar qq)_{V-A} \\
Q_4 &=& (\bar s_i b_j)_{V-A} \sum_q (\bar q_j q_i)_{V-A} \\
Q_5 &=& (\bar sb)_{V-A} \sum_q (\bar qq)_{V+A} \\
Q_6 &=& (\bar s_i b_j)_{V-A} \sum_q (\bar q_j q_i)_{V+A} \\
\label{q7def}
Q_7 &=& \frac{e}{8\pi^2}m_b\, 
        \bar s_i\sigma^{\mu\nu}(1+\gamma_5)b_i\, F_{\mu\nu}\\
\label{q8def}
Q_8 &=& \frac{g}{8\pi^2}m_b\, 
        \bar s_i\sigma^{\mu\nu}(1+\gamma_5)T^a_{ij} b_j\, G^a_{\mu\nu}
\end{eqnarray}
The most important operators are the magnetic penguin operator $Q_7$ and the four-quark operators $Q^p_{1,2}$. The sign conventions for the electromagnetic and strong couplings correspond to the covariant derivative $D_\mu=\partial_\mu +ie Q_f A_\mu + i g T^a A^a_\mu$. With these definitions the coefficients $C_{7,8}$ are negative in the Standard Model, which is the choice generally adopted in the literature. The effective Hamiltonian for $b\to d\gamma\gamma$ is obtained from (\ref{heff}--\ref{q8def}) by the replacement $s\to d$. The Wilson coefficients $C_i$ in (\ref{heff}) are known at next-to-leading order \cite{CMM}.

The amplitude for the $B\to\gamma\gamma$ decay has the general structure
\begin{eqnarray}\label{Bgggen}
\lefteqn{{\cal A}
\left(\bar B\to\gamma(k_1,\epsilon_1)\gamma(k_2,\epsilon_2)\right)\equiv
 \frac{G_F}{\sqrt{2}} \frac{\alpha}{3\pi} f_B 
 \frac{1}{2}\langle\gamma\gamma|
 A_+ F_{\mu\nu} F^{\mu\nu}-i\, A_- F_{\mu\nu} \tilde F^{\mu\nu}|0\rangle 
  } \nonumber\\
  &&= \frac{G_F}{\sqrt{2}} \frac{\alpha}{3\pi} f_B \left[
 A_+\left(
  2k_1\cdot\epsilon_2 \, k_2\cdot\epsilon_1 -m_B^2 \epsilon_1\cdot\epsilon_2
 \right) -2i\,A_-\varepsilon(k_1,k_2,\epsilon_1,\epsilon_2)\right]
\end{eqnarray}
Here $F^{\mu\nu}$ and $\tilde F^{\mu\nu}$ are the photon field strength 
tensor and its dual, where
\begin{equation}\label{fdual}
\tilde F^{\mu\nu}=\frac{1}{2}\varepsilon^{\mu\nu\lambda\rho}F_{\lambda\rho}
\end{equation}
with $\varepsilon^{0123}=-1$.
The decay rate is then given by
\begin{equation}\label{gbgg}
\Gamma(\bar B\to\gamma\gamma)=
\frac{G^2_F m^3_B f^2_B \alpha^2}{288\pi^3}\left(|A_+|^2 + |A_-|^2\right)
\end{equation}

In the heavy-quark limit we propose a factorization formula for the hadronic 
matrix elements of the operators in (\ref{heff}):
\begin{equation}\label{fform}
\langle \gamma(\epsilon_1)\gamma(\epsilon_2)|Q_i|\bar B\rangle = f_B
\int^1_0 d\xi\, T^{\mu\nu}_i(\xi)\, \Phi_B(\xi)\epsilon_{1\mu} \epsilon_{2\nu}
\end{equation}
where the $\epsilon_i$ are the polarization 4-vectors of the photons and 
$\Phi_B\equiv\Phi_{B1}$ is the leading twist light-cone distribution 
amplitude of the $B$ meson. 
The latter quantity is a universal, nonperturbative object.
It is defined via the light-cone projector for the $B$ meson
at leading power \cite{BBNS}
\begin{equation}\label{phi12def}
  \langle 0|b(0) \bar q(z)|\bar B(p)\rangle = \frac{i f_B}{4}(\not\! p + m_b)\gamma_5\, \int^1_0 d\xi\, e^{-i\xi p_+ z_-}[\Phi_{B1}(\xi)+\not\! n \Phi_{B2}(\xi)]
\end{equation}
The functions $\Phi_{B1,B2}(\xi)$ describe the distribution of 
light-cone momentum fraction $\xi=l_+/p_+$ of the spectator quark $q$ with 
momentum $l$ inside the $B$ meson (only $\Phi_{B1}$ contributes in
\ref{fform}). 
Here light-cone components of four-vectors $v$ are defined by
\begin{equation}\label{vlcc}
  v_\pm =\frac{v^0\pm v^3}{\sqrt{2}}
\end{equation}
The wave functions are highly asymmetric with 
$\xi={\cal O}(\Lambda_{QCD}/m_b)$. They are normalized as
\begin{equation}\label{phi12norm}
  \int^1_0 d\xi\, \Phi_{B1}(\xi)=1 \qquad  \int^1_0 d\xi\, \Phi_{B2}(\xi)=0
\end{equation}
If the light-like vector $n$ in (\ref{phi12def}) is chosen appropriately 
parallel to one of the 4-momenta of the photons, only the first negative 
moment of $\Phi_{B1}(\xi)$ appears, which we parametrize by a quantity 
$\lambda_B={\cal O}(\Lambda_{QCD})$, i.e.
\begin{equation}\label{lambdef}
  \int^1_0 d\xi \frac{\Phi_{B1}(\xi)}{\xi}=\frac{m_B}{\lambda_B}
\end{equation}
Because there are no hadrons in the final state, only one type  
of hard-scattering kernel $T$ 
(type II or hard-spectator contribution \cite{BB,BBNS})
enters the factorization 
formula. The QCD factorization formula (\ref{fform}) holds up to corrections 
of relative order $\Lambda_{QCD}/m_b$.
The form of (\ref{fform}) with a simple convolution over the light-cone
variable $\xi$ is appropriate for the lowest (leading logarithmic)
order in $\alpha_s$, which we will use in the present analysis.
A generalization to include transverse-momentum variables is likely to
be necessary at higher orders in QCD \cite{BBNS,KPY,BF,BFPS}.

We conclude this section with a brief discussion of
CP violation in $B\to\gamma\gamma$.
The subscripts $\pm$ on $A_\pm$, 
defined in (\ref{Bgggen}) for $\bar B\to\gamma\gamma$, 
denote the CP properties of the corresponding two-photon final states,
which are eigenstates of CP: 
$A_{CP=+1}$ is proportional to the 
parallel spin polarization $\vec\epsilon_1\cdot\vec\epsilon_2$ and 
$A_{CP=-1}$ is proportional to the perpendicular spin polarization 
$\vec\epsilon_1\times\vec\epsilon_2$ of the photons. 
In addition to $A_\pm$ we introduce the
CP conjugated amplitudes $\bar A_\pm$ for the decay 
$B\to\gamma\gamma$ (decaying $\bar b$ anti-quark).
Then the deviation of the ratios
\begin{equation}\label{rCPdef}
  r^\pm_{CP}=\frac{|A_\pm|^2-|\bar A_\pm|^2}{|A_\pm|^2+|\bar A_\pm|^2}
\end{equation}
from zero is a measure of direct CP violation. We shall
focus here on this direct effect, which is specific for the
$B\to\gamma\gamma$ decay, and will not consider
CP violation originating in $B$--$\bar B$ mixing.

Due to the unitarity of the CKM matrix we can parametrize
\begin{eqnarray}\label{apar}
  A_\pm &=& \lambda_u a_u^\pm e^{i\alpha_u^\pm} +\lambda_c a_c^\pm e^{i\alpha_c^\pm}\\ \label{abarpar}
  \bar A_\pm &=& \lambda_u^* a_u^\pm e^{i\alpha_u^\pm} +\lambda_c^* a_c^\pm e^{i\alpha_c^\pm}
\end{eqnarray}
where $a_{u,c}^\pm$ are real hadronic matrix elements of weak transition 
operators and the $\alpha_{u,c}^\pm$ are their CP-conserving phases. 
We then have
\begin{equation}\label{rCPpar}
  r^\pm_{CP}=\frac{2 \,\mathrm{Im}(\lambda_u \lambda_c^*)a_u^\pm a_c^\pm \sin(\alpha_c^\pm-\alpha_u^\pm)}{|\lambda_u|^2 a_u^{\pm 2} +|\lambda_c|^2 a_c^{\pm 2} +2 \,\mathrm{Re}(\lambda_u\lambda_c^*)a_u^\pm a_c^\pm \cos(\alpha_c^\pm -\alpha_u^\pm)}
\end{equation}
The weak phase differences read
\begin{equation}
  \mathrm{Im}(\lambda_u^{(q)} \lambda_c^{(q)*})=
    \mp|\lambda_u^{(q)}||\lambda_c^{(q)}|\sin\gamma
\end{equation}
with the negative sign for $q=s$ and the positive sign for $q=d$.
The strong phase differences $\sin(\alpha_c^\pm -\alpha_u^\pm)$
arise from final state interaction (FSI) effects,
generated for instance via the Bander-Silverman-Soni (BSS) mechanism 
\cite{BSS}.

\section{$B\to\gamma\gamma$ at leading power}\label{sec:ldgpow}

\begin{figure}[t]
\begin{center}
\psfig{figure=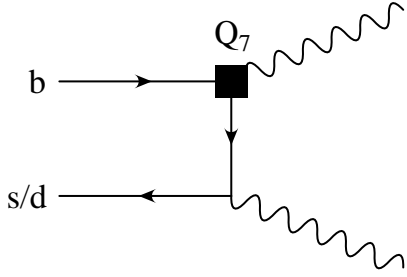}
\end{center}
\caption{The leading power contribution to the $B\to\gamma\gamma$ amplitude 
given by the magnetic penguin operator $Q_7$. The diagram with interchanged 
photons is not shown. Radiation from the $b$-quark line 
(Fig.~\ref{fig:1PRslp}) is power suppressed. 
\label{fig:1PRlp}}
\end{figure}

To leading power in $\Lambda_{QCD}/m_b$ and
in the leading logarithmic (LL) approximation of QCD only one diagram 
contributes to the amplitude for $B\to\gamma\gamma$. It is the one-particle 
reducible (1PR) diagram with the electromagnetic penguin operator $Q_7$, 
where the second photon is emitted from the $s$-quark line. 
An illustration is given in Fig.~\ref{fig:1PRlp}.
Evaluating the factorization formula (\ref{fform}) for this diagram leads 
to the matrix element
\begin{eqnarray}\label{Q7lp}
\lefteqn{\langle\gamma(k_1,\epsilon_1)\gamma(k_2,\epsilon_2)|Q_7|
\bar B\rangle =}\\
&& i\frac{f_B \alpha}{\pi} Q_s \int^1_0 d\xi\, \frac{\Phi_{B1}(\xi)}{\xi} 
\left(2k_1\cdot\epsilon_2 k_2\cdot\epsilon_1-
m_B^2\epsilon_1\cdot\epsilon_2 -
2i\,\varepsilon(k_1,k_2,\epsilon_1,\epsilon_2)
\right)\nonumber
\end{eqnarray}
Together with (\ref{heff}) and the definition of $A_\pm$
in (\ref{Bgggen}) this implies 
\begin{equation}\label{apmlp}
A_\pm=\lambda^{(q)}_t C_7 \frac{m_{B}}{\lambda_{B}}
\end{equation}
where CKM unitarity, $\lambda^{(q)}_u+\lambda^{(q)}_c=-\lambda^{(q)}_t$,
has been used.
After summing over the photon polarizations  
the branching fraction to leading power becomes
\begin{equation}\label{brlp}
B(B_q\to\gamma\gamma)=\tau_{B_q} \frac{G^2_F m^5_{B_q}}{192\pi^3}
|\lambda^{(q)}_t|^2 C^2_7 \frac{4\alpha^2 f^2_{B_q}}{3\lambda^2_{B_q}}
\end{equation}
where $q=d,s$ for the decay of a $B_d$ or a $B_s$ meson, respectively.

In the present approximation the strong-interaction matrix elements
multiplying $\lambda^{(q)}_u$ and $\lambda^{(q)}_c$ are identical
and have no relative phase.
The CP-violating quantities $r^\pm_{CP}$ defined
in (\ref{rCPdef}) therefore vanish for the strictly leading-power result.

One may compare the behaviour of the $B\to\gamma\gamma$ 
decay amplitude in the heavy-quark limit with that of the rare decays 
$B\to V\gamma$ (with $V$ a vector meson) and $B\to\pi\pi$. 
Disregarding CKM factors we can write
\begin{equation}\label{ggvgpp}
\begin{array}{lll}
{\cal A}(B\to\gamma\gamma) & \sim  G_F\ e^2\ f_B A_\pm m^2_B
                            & \sim   G_F\ e^2\ \Lambda^{1/2} m^{5/2}_b\\
{\cal A}(B\to V\gamma) &\sim  G_F\ e\ m_b F_V m^2_B
                             &\sim   G_F\ e\ \Lambda^{3/2} m^{3/2}_b\\
{\cal A}(B\to\pi\pi) &\sim  G_F\ e^0\ f_\pi F^{B\to\pi} m^2_B
                             &\sim   G_F\ e^0\ \Lambda^{5/2} m^{1/2}_b
\end{array}
\end{equation}
where we have used that the form factors scale as ($\Lambda=\Lambda_{QCD}$)
\begin{equation}
A_\pm\sim \frac{m_b}{\Lambda}\qquad f_B\sim\frac{\Lambda^{3/2}}{m^{1/2}_b}
\qquad F_V\sim F^{B\to\pi}\sim\frac{\Lambda^{3/2}}{m^{3/2}_b}
\end{equation}
We observe that the ratio of the three amplitudes in (\ref{ggvgpp})
is $1:(\Lambda/m_b):(\Lambda/m_b)^2$. Thus ${\cal A}(B\to\gamma\gamma)$
is leading in the heavy-quark limit, while the other two are successively
stronger suppressed. However, this hierarchy is compensated by the opposite 
pattern $e^2:e:1$ in the electromagnetic coupling $e$.

\section{Contributions to $B\to\gamma\gamma$ with power suppression}
\label{sec:subldgpow}

\begin{figure}[t]
\begin{center}
\psfig{figure=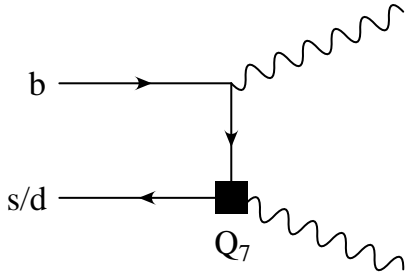}
\end{center}
\caption{The subleading power 1PR diagram of the magnetic penguin operator 
$Q_7$. The diagram with interchanged photons is not shown.\label{fig:1PRslp}}
\end{figure}

\begin{figure}[t]
\begin{center}
\psfig{figure=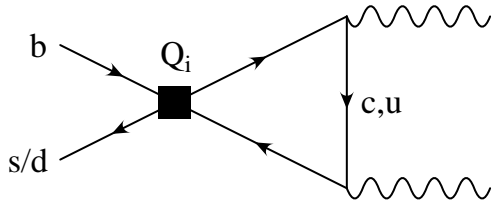}
\end{center}
\caption{The subleading power 1PI diagram. Here the four-quark operators 
$Q_i$ can contribute. The diagram with interchanged photons is not shown.
\label{fig:1PI}}
\end{figure}

Subleading contributions come from the 1PR diagram, where the second photon 
is emitted off the $b$ quark line (Fig. \ref{fig:1PRslp}), and from the 
one-particle irreducible diagram (1PI) (see Fig. \ref{fig:1PI}).

To estimate the subleading 1PR contribution, we simply evaluate
the graphs in Fig. \ref{fig:1PRslp} using the $B$-meson projector
in (\ref{phi12def}).
The result equals $\lambda_B/m_B$ times the leading-power expression
in (\ref{Q7lp}), clearly showing the power suppression of this mechanism.
Other corrections of the same order can arise from higher-twist terms in 
the $B$-meson wave function. We shall neglect all those subleading terms
for the matrix element of $Q_7$, keeping in mind that
they could naturally contribute relative corrections of order $\sim 10\%$.
We stress, however, that subleading effects in $\langle Q_7\rangle$
contribute equally to the up- and charm-quark components of the amplitude
(see (\ref{heff})). Therefore they do not give rise, in particular,
to relative FSI phases between these sectors, which would affect direct 
CP violation. The same is true for perturbative QCD corrections to
$\langle Q_7\rangle$.

We next consider more closely the diagrams in Fig. \ref{fig:1PI}.
These 1PI contributions, which come from the matrix elements of
four-quark operators in (\ref{heff}), are of special interest for
two reasons.
First, they are the basic effects responsible for a difference
between the up- and charm-quark sectors of the amplitude, including
rescattering phases. Second, they represent the parton-level
processes that are dual to $B\to\gamma\gamma$ amplitudes from
$Q_i$ with hadronic intermediate states ($J/\Psi$, $D^{(*)}$, etc.),
which are commonly considered as generic long-distance contributions.
Investigating the amplitude in Fig. \ref{fig:1PI} will therefore
shed light on this class of effects in $B\to\gamma\gamma$.

\begin{figure}[t]
\psfig{figure=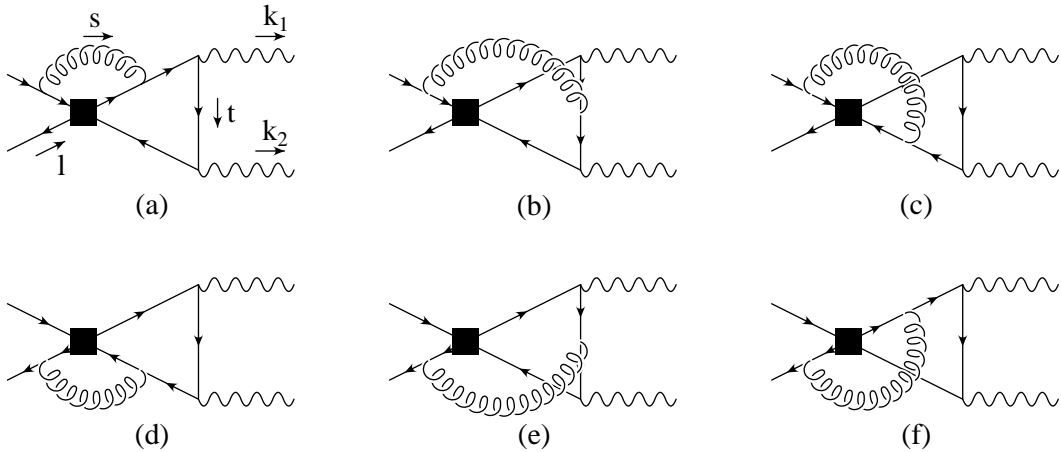}
\caption{Gluon corrections to the 1PI diagram. Quark labels are as in Fig.~\ref{fig:1PI} and the momentum assignment is everywhere as in diagram (a). The diagrams with interchanged photons are not shown.}
\label{fig:1PINLO}
\end{figure}

It shall now be argued that the 1PI contributions are
calculable using QCD factorization. 
It is known that the one-loop contribution in Fig. \ref{fig:1PI} is
free of infrared (IR) singularities. In addition we will show that also at
${\cal O}(\alpha_s)$ there are no collinear or soft 
IR divergences at leading power in $1/m_b$. 
By leading power we here mean the lowest nonvanishing order in the
power expansion, where the entire 1PI contribution starts only at
subleading power with respect to the dominant mechanism in
Fig. \ref{fig:1PRlp}. This suppression will be apparent from the explicit
expressions given in (\ref{1PIQ12}) below.
The relevant two-loop diagrams at ${\cal O}(\alpha_s)$ are shown
in Fig.~\ref{fig:1PINLO}.
Following the approach  explained in the second reference
of \cite{BBNS} for the case of $B\to D\pi$ decays,
we demonstrate the absence of divergences for the 
hard-scattering part of the amplitude at ${\cal O}(\alpha_s)$.   
As a consequence the matrix elements of four-quark operators
can be written in factorized form as in (\ref{fform}),
where in the present case the kernel $T^{\mu\nu}_i$ is independent
of $\xi$:
\begin{equation}\label{fform12}
\langle \gamma(\epsilon_1)\gamma(\epsilon_2)|Q_{1,2}|\bar B\rangle = 
f_B T^{\mu\nu}_{1,2}\, \epsilon_{1\mu} \epsilon_{2\nu}
\end{equation}

The proof proceeds by identifying the potentially IR singular
regions in the loop-momentum variables $s$ and $t$ and by determining the
degree of IR divergence through power counting. We distinguish the
following cases: (i) $s$, $t$ soft, where all components scale like
$s$, $t\sim\Lambda\ll m_b$.
(ii) $s$, $t$ collinear with either photon momentum $k_1$ or $k_2$,
for instance
\begin{equation}\label{stcoll}
s=\alpha k_1+s_\perp +\alpha_2 k_2\qquad\qquad
t=\beta k_1 + t_\perp + \beta_2 k_2
\end{equation}
describing the region where $s$ and $t$ are collinear with $k_1$.
Similar relations, with $k_1$ and $k_2$ interchanged, apply to the
case where the loop momenta become collinear with $k_2$.
The parameters in (\ref{stcoll}) scale as
\begin{equation}\label{albescale}
\alpha, \beta\sim 1, \qquad s_\perp, t_\perp\sim\Lambda, \qquad 
\alpha_2, \beta_2\sim \frac{\Lambda^2}{m^2_b}
\end{equation}
 and we have
\begin{equation}\label{k12stperp}
k^2_{1,2}=k_{1,2}\cdot s_\perp=k_{1,2}\cdot t_\perp=0 
\end{equation}
We consider explicitly the case where the quark inside the loop
is massless, which is the most IR singular situation. The momentum $l$ of
the light external quark is counted as a soft quantity $l\sim\Lambda$.
The dominant region for the diagrams in Figs. \ref{fig:1PI}, 
\ref{fig:1PINLO} arises when both loop momenta are hard, $s$, $t\sim m_b$.
The diagrams then scale as $m_b$. To demonstrate IR finiteness, one has to
show that the potentially singular regions scale at least with one power
of $\Lambda$ and thus are suppressed relative to the hard contribution.
Scaling as $\Lambda^0 m_b$ in a singular region would indicate a
logarithmic divergence.
The relevant cases for $s$ and $t$ are the soft-soft, soft-hard,
hard-soft, collinear-collinear, collinear-hard, hard-collinear,
soft-collinear and collinear-soft regions.
For the collinear regions the momenta can be either $\sim k_1$ or
$\sim k_2$. In the collinear-collinear case one needs to consider
primarily that both momenta $\sim k_1$, or both $\sim k_2$. If one momentum
is $\sim k_1$ and the other $\sim k_2$, the contribution is always
less singular in the infrared.
In the second reference of \cite{BBNS}, besides the soft-collinear
regions, a supersoft-collinear scaling had to be discussed,
where $s$, $t\sim\Lambda^2$ for supersoft momenta. In the
present context this scaling is not more singular than the 
soft-collinear one and will thus not be considered further.

Power counting then shows that all diagrams in Fig. \ref{fig:1PINLO} are 
IR finite at leading power
in the soft-soft, soft-hard, hard-soft, collinear-hard, hard-collinear,
soft-collinear and collinear-soft regions. In some cases one encounters
a superficial logarithmic divergence, which vanishes when the relations
$k^2_{1,2}=0$ and $k_1\cdot\epsilon_1=k_2\cdot\epsilon_2=0$ are used.

On the other hand, in the collinear-collinear region (superficial)
linear divergences $\sim\Lambda^{-1}$ may appear. An example is
diagram (f) in Fig. \ref{fig:1PINLO} when both $s$ and $t$ are collinear
to $k_1$.
The diagram has the form
\begin{equation}\label{diagf}
D_{(f)}=d^4s\, d^4t
\frac{\gamma_\sigma(\not\! l-\not\!s)\gamma^\lambda(1-\gamma_5)
(\not\! t-\not\! k_2)\gamma_\nu\not\! t\gamma_\mu(\not\! t+\not\! k_1)
\gamma^\sigma(\not\! t+\not\! k_1-\not\! s)\gamma_\lambda(1-\gamma_5)}{
s^2 \,(l-s)^2\, (t-k_2)^2\, t^2\, (t+k_1)^2\, (t+k_1-s)^2}
\end{equation}
Expanding the numerator in powers of $\Lambda$ using (\ref{stcoll})
one finds that the superficial linear divergence drops out.
This leads to
\begin{equation}\label{diagfk1}
D_{(f)}=d^4s\, d^4t\frac{\gamma_\sigma (\not\! l-\not\! s)}{(l-s)^2}
\Gamma_R\, k^\sigma_1
\end{equation}
with the remainder of the diagram $\Gamma_R\sim\Lambda^{-7}$.
Together with collinear phase space, $d^4s\, d^4 t\sim\Lambda^8$,
the integrand then appears to scale as $\Lambda^0 m_b$.
In fact, since $\not\! k_1(\not\! l-\not\! s)\sim m_b\Lambda$,
$(l-s)^2\sim m_b \Lambda$ for $s=\alpha k_1+{\cal O}(\Lambda)$,
the contribution behaves only as $D_{(f)}\sim \Lambda$.
In a similar way all other contributions from the
collinear-collinear regions can be shown to be power-suppressed.

In addition to the graphs displayed in Fig. \ref{fig:1PINLO}
there are also diagrams in which the gluon connects only to the lines
inside the triangular quark loop (vertex and self-energy corrections).
Also these diagrams are found to be IR finite at leading power.

Finally, in contrast, one-gluon exchange among the external lines 
does generate leading IR divergences. However, these singularities
match those appearing in the perturbative evaluation of the
$B$-meson decay constant $f_B$ at the same order, in accordance
with the factorization formula (\ref{fform12}).
This completes our proof that at ${\cal O}(\alpha_s)$ for the 1PI 
contribution there are no collinear or soft infrared divergences at leading 
power in $1/m_b$ in the hard-scattering kernel. 
It supports our claim that these diagrams are calculable using QCD 
factorization. 

\begin{figure}[t]
   \epsfxsize=12cm
   \centerline{\epsffile{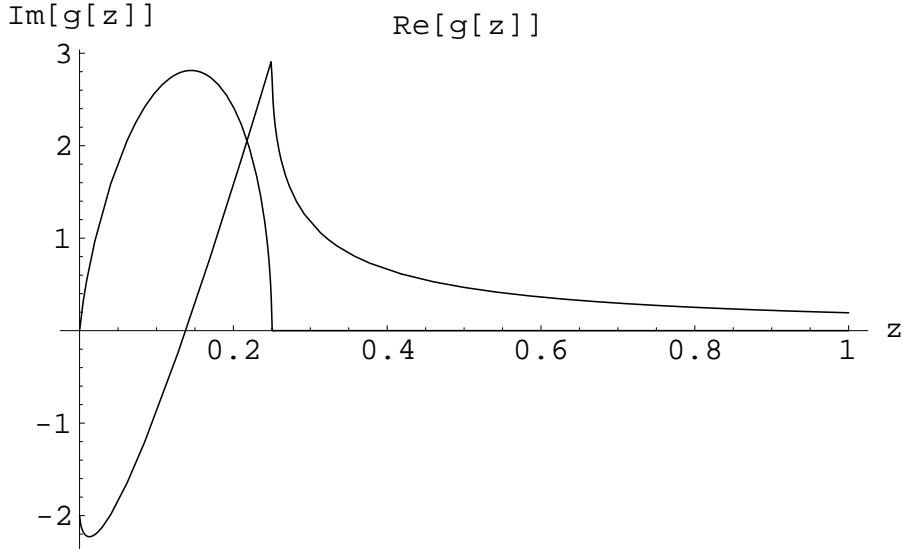}}
\caption{Real and imaginary part of $g(z)$.\label{fig:ggzplot}}
\end{figure}

Evaluating the 1PI diagram in Fig.~\ref{fig:1PI} we find
the matrix elements
\begin{eqnarray}\label{1PIQ12}
  \left.\begin{array}{ll} \langle Q^p_1\rangle\\ \langle Q^p_2\rangle 
        \end{array}\right\} &=& 
       -\frac{f_{B_q}\alpha}{\pi} Q_u^2 g(z_p) 
    \varepsilon(k_1,k_2,\epsilon_1,\epsilon_2) \left\{\begin{array}{cc} 1\\ N 
\end{array}\right.
\end{eqnarray}
with
\begin{equation}\label{jofz}
  g(z)=-2+4z\left[L_2\left(\frac{2}{1-\sqrt{1-4z+i\epsilon}}\right)+
       L_2\left(\frac{2}{1+\sqrt{1-4z+i\epsilon}}\right)\right]
\end{equation}
Here
\begin{equation}\label{dilog}
  L_2(x)=-\int^x_0 dt\frac{\ln(1-t)}{t}
\end{equation}
and ($p=u,c$)
\begin{equation}\label{zdef}
  z_p=\frac{m_p^2}{m_b^2}
\end{equation}
We note that the function $g(z)$ is related to the
kernel $h(u,s)$ defined in \cite{BB} in the context of $B\to V\gamma$
by $g(z)=h(1,z)$. Real and imaginary part of $g(z)$ are
displayed in Fig. \ref{fig:ggzplot}.
The function $g(z)$ is regular for $z\to 0$, $g(0)=-2$. Its value at 
$z=1$ is $g(1)=2(\pi^2-9)/9$, and $g(z)\to 0$ as $z\to\infty$.
The function $g(z)$ has an 
imaginary part for $0 < z < 1/4$, which is maximal close to
$z\approx 0.1$, the approximate physical value for the case
of an internal charm quark.
The imaginary part is due to the rescattering process
$B\to p\bar p\to\gamma\gamma$ with an on-shell light-quark pair $p\bar p$
in the intermediate state. Since the process $B\to p\bar p$ is
forbidden by helicity and angular momentum conservation for a
massless quark $p$, the function $g(z)$ is real at $z=0$ (that is
for $p=u$). The helicity suppression of the phase will be absent at higher
order in $\alpha_s$. 

Including the effect of 1PI diagrams in the decay amplitudes,
the quantities $A_\pm$ from (\ref{apmlp}) become
\begin{eqnarray}
A^p_+ &=& -C_7 \frac{m_{B}}{\lambda_{B}} \label{apslp} \\
A^p_- &=& -C_7 \frac{m_{B}}{\lambda_{B}}-\frac{2}{3}(C_1+N C_2)g(z_p)
\label{amslp}
\end{eqnarray}
where we have decomposed $A_\pm$ into
\begin{equation}\label{apmap}
A_\pm=\sum_{p=u,c} \lambda^{(q)}_p A^p_\pm
\end{equation}
We see that only $A_-$ is modified. The imaginary part of the 1PI loop
diagram leads to a relative CP-conserving phase between the $\lambda_u$ and 
the $\lambda_c$ contribution. This gives a nonvanishing 
$r^-_{CP}$ while $r^+_{CP}$ is still zero.

The effect of  penguin operators $Q_3,\ldots, Q_6$ in the 1PI
graphs is very small (at the percent level for the
total amplitude) due to the small size of
their Wilson coefficients. Also, they contribute equally to the
up and charm sectors. We shall neglect them here. Their matrix elements
can be found in \cite{SWB}.


\section{Long-distance contributions to $B\to\gamma\gamma$ and
         $D\to\gamma\gamma$}
\label{sec:ldcont}

We would next like to comment on the issue
of long-distance effects and their relation with 
our analysis based on the heavy-quark limit.
Long-distance contributions were calculated for both
$B\to\gamma\gamma$ \cite{HI,CE,LZZ} and $D\to\gamma\gamma$ \cite{FSZ,BGHP}.
The following mechanisms have been considered:

\begin{itemize}

\item
$B$ and $D\to V\gamma$, with a light vector meson $V$, followed
by a $V\to\gamma$ conversion with the conversion factor supplied by
the vector meson dominance (VMD) model. This gives a long-distance
contribution to the matrix element of operator $Q_7$. Counting powers
of $m_b$ in the explicit result for the $B_s\to\phi\gamma\to\gamma\gamma$
amplitude in \cite{HI} shows that it is suppressed with
$\Lambda_{QCD}/m_b$ compared to the leading power result from
Fig.~\ref{fig:1PRlp}. However, this is only one among the
possible subleading power mechanisms for $\langle Q_7\rangle$.
Another comes, for example, from the diagram in Fig.~\ref{fig:1PRslp}.

We can apply a similar power counting to the $D\to\gamma\gamma$
VMD amplitude of \cite{BGHP}. In the limit of a heavy charm quark it
is likewise power suppressed with $\Lambda_{QCD}/m_c$ compared to
the short-distance amplitude. The limit $m_c\gg\Lambda_{QCD}$ is
certainly questionable, but has nevertheless proven successful
in some applications to charmed hadrons \cite{MN}. 
Again, the VMD contributions are just one out of several
possible power corrections to the $Q_7$ matrix element.

\item
Single-particle unitarity contributions where the $D^0$ mixes with
a spinless intermediate particle $M$, which then decays into a
photon pair, $D^0\to M\to\gamma\gamma$. As far as the quark-level
topology is concerned, such a process corresponds to the 1PI
contribution in Fig.~\ref{fig:1PI}. Relative to this contribution the
amplitudes in \cite{BGHP} are at most of the same order in powers of the 
inverse heavy charm quark mass.

\item
Two-particle unitarity contributions are other hadron-level
counterparts of the 1PI diagrams. Triangle loops with intermediate
$D^{(*)}_s$ \cite{CE,LZZ} and $K^{(*)}$ \cite{BGHP} mesons were
calculated for $B_s\to\gamma\gamma$ and $D\to\gamma\gamma$, respectively.
As we will see in the next section, the effect of the 1PI contributions
in $B\to\gamma\gamma$ on the decay amplitude is small. The authors of
\cite{LZZ} also get a not too large effect in their reestimate of
the result of \cite{CE}, which had suggested a more sizable long-distance
contribution.
Yet, one has to keep in mind that the effects from intermediate $D$ and
$D^*$ mesons are only two of many possible hadronic intermediate states. 
Since the meson model calculations are not based on a systematic
approximation, it is for instance conceivable that cancellations
required by quark-hadron duality would be missed. This could then
lead to an overestimate of the effect.

With obvious replacements we can use our 1PI results in
(\ref{1PIQ12}) - (\ref{apmap}) to estimate the quark-level analogue
of the two-particle unitarity contributions to $D\to\gamma\gamma$
calculated in \cite{BGHP}. We obtain for the 1PI amplitude

\begin{eqnarray}\label{adgg1pi}
A^{D\to\gamma\gamma}_{\rm{1PI}} &=&
  \frac{G_F}{\sqrt{2}}\frac{\alpha}{\pi}f_D Q^2_d\, (C_1+N C_2)
 |V_{cs}| |V_{us}| \nonumber\\
&& \cdot\left[ 
 g\left(\frac{m^2_s}{m^2_c}\right)-g\left(\frac{m^2_d}{m^2_c}\right) \right]
 \, i\varepsilon(\epsilon_1,\epsilon_2,k_1,k_2) \nonumber\\
&\approx& -2.4\cdot 10^{-11}{\rm{GeV}}^{-1}\, [-0.2+0.5i]
   \, \varepsilon(\epsilon_1,\epsilon_2,k_1,k_2) \nonumber\\
\left|A^{D\to\gamma\gamma}_{\rm{1PI}}\right| &\approx&
 1.3\cdot 10^{-11} {\rm{GeV}}^{-1} \,
  |\varepsilon(\epsilon_1,\epsilon_2,k_1,k_2)|
\end{eqnarray}
where we set $f_D=200\,\rm{MeV}$, $m_s=120\,\rm{MeV}$, $m_d=0$ 
and evaluated the Wilson coefficients
at $m_c$. The contributions of a strange quark and a down quark in the loop
are accompanied by CKM elements that are practically equal in absolute value,
but opposite in sign. This leads to a strong GIM cancellation in the
square bracket.  
Note that
\begin{equation}\label{ggsmallz}
g(z)=-2+(-2\ln^2z+2\pi^2 -4\pi\, i\, \ln z)z+ {\cal O}(z^2)
\end{equation}
The GIM mechanism removes the constant $(-2)$ and gives an additional
suppression $\sim 1/m^2_c$. Due to large logarithms $\ln z$ and
numerical factors the suppression is relatively mild numerically.
Also for the 1PR amplitude, the $c\to u\gamma$ 
transition exhibits a much stronger GIM cancellation than the corresponding
process $b\to s\gamma$. In this case this has crucial consequences 
for the hierarchy
of higher-order QCD terms. Leading logarithmic QCD corrections are known 
to enhance the $c\to\ u \gamma$ amplitude by more than an order of
magnitude. Including the two-loop QCD contributions increases
the amplitude by another two orders of magnitude \cite{GHMW}.
Note however that the two-loop estimate
from \cite{GHMW} shows a GIM cancellation pattern very similar to
the 1PI amplitude in (\ref{adgg1pi}). In other words, the
strong hierarchy of QCD corrections discussed in \cite{GHMW} is
peculiar to the $c\to u\gamma$ transition. Hence, we do not expect
a significant change from higher-order QCD corrections to (\ref{adgg1pi}).
As a crude estimate, we therefore compare (\ref{adgg1pi}) with the
$c\to u\gamma$ amplitude from the dominant two-loop diagrams
of \cite{GHMW}, which lead to an effective coefficient
$|A|=0.0047$ (entering the amplitude with a normalization
corresponding to $|V_{cs}V_{us} C_7|$). 
We thus obtain for the ``short-distance'' 
amplitude (assuming $\lambda_D=350\,\rm{MeV}$):
\begin{eqnarray}\label{adggsd}
A^{D\to\gamma\gamma}_{\rm{SD}} &=&
 -\frac{G_F}{\sqrt{2}}\frac{\alpha}{\pi}f_D\, 2Q_u\, A\frac{m_D}{\lambda_D}
   \nonumber\\
&&\cdot\left[ i\varepsilon(\epsilon_1,\epsilon_2,k_1,k_2) 
       -k_1\cdot\epsilon_2\, k_2\cdot\epsilon_1+\frac{m^2_D}{2}
        \epsilon_1\cdot\epsilon_2\right]   \nonumber\\
\left|A^{D\to\gamma\gamma}_{\rm{SD,-}}\right| &\approx&
12.8\cdot 10^{-11} {\rm{GeV}}^{-1} |\varepsilon(\epsilon_1,\epsilon_2,k_1,k_2)|
\end{eqnarray}
We see that the 1PI amplitude, in this rough order-of-magnitude estimate,
is smaller than the CP-odd part of the short-distance amplitude. 

The authors of \cite{BGHP}, on the other hand, find a large effect
on the $D\to\gamma\gamma$ branching ratio from $K^+K^-$
intermediate states. Their amplitude, however, is formally
suppressed with $1/m_c$ relative to our 1PI
amplitude. In view of our estimate above the large effect could
be overrated. Only one intermediate state was taken into account so that
important contributions in the duality sum might be missing. 

\item
The decay mechanism $B_s\to\phi\psi\to\phi\gamma\to\gamma\gamma$
was estimated to give very small effects on the $B\to\gamma\gamma$
decay amplitudes \cite{HI}. It is power suppressed compared to the
leading 1PR contribution.
\end{itemize}

\section{Phenomenology}\label{sec:phen}
In this section we present numerical results for the expressions derived 
in this paper. Our choice of input parameters is summarized in 
Table~\ref{tab:input}.
\begin{table}[tpb]
\renewcommand{\arraystretch}{1.2}
\begin{center}
\begin{tabular*}{140mm}{@{\extracolsep\fill}|c|c|c|c|c|c|}
\hline\hline
\multicolumn{6}{|c|}{CKM parameters and coupling constants}\\
\hline
$V_{us}$ & $V_{cb}$ & $\left|V_{ub}/V_{cb}\right|$ & 
$\Lambda_{\overline{MS}}^{(5)}$ & $\alpha$ & $G_F$\\
\hline
0.22 & 0.041 & $0.085 \pm 0.025$ & 225 MeV & 1/137 & $1.166 \times 10^{-5} 
\,\mbox{GeV}^{-2}$\\
\hline
\end{tabular*}

\begin{tabular*}{140mm}{@{\extracolsep\fill}|c|c|c|c|}
\hline
\multicolumn{4}{|c|}{Parameters related to the $B_d$ meson}\\
\hline
$m_{B_d}$ & $f_{B_d}$ \cite{SMR}& $\tau_{B_d}$ & $\lambda_{B_d}$\\
\hline
5.28 GeV & (200 $\pm$ 30) MeV & 1.55 ps & $(350 \pm 150)$ MeV\\
\hline
\end{tabular*}

\begin{tabular*}{140mm}{@{\extracolsep\fill}|c|c|c|c|}
\hline
\multicolumn{4}{|c|}{Parameters related to the $B_s$ meson}\\
\hline
$m_{B_s}$ & $f_{B_s}$ \cite{SMR}& $\tau_{B_s}$ & $\lambda_{B_s}$\\
\hline
5.37 GeV & (230 $\pm$ 30) MeV & 1.49 ps & $(350 \pm 150)$ MeV\\
\hline
\end{tabular*}

\begin{tabular*}{140mm}{@{\extracolsep\fill}|c|c|c|c|}
\hline
\multicolumn{4}{|c|}{Quark and W-boson masses}\\
\hline
$m_b(m_b)$ & $m_c(m_b)$ & $\overline{m}_{t}(m_t)$ & $M_W$\\
\hline
4.2 GeV & $(1.3 \pm 0.2)$ GeV & 166 GeV & 80.4 GeV\\
\hline\hline
\end{tabular*}
\end{center}
\caption[]{Summary of input parameters.\label{tab:input}}
\end{table}
For definiteness we employ the two-loop form of the running coupling
$\alpha_s(\mu)$ (as quoted in \cite{BBL}), which corresponds to
$\alpha_s(M_Z)=0.118$ for $\Lambda^{(5)}_{\overline{MS}}=225\,{\rm MeV}$.
With central values of all input parameters, at $\mu=m_b$, and using
a nominal value of the CKM angle $\gamma=58^\circ$, we find for 
the branching ratios to leading logarithmic accuracy, 
evaluating (\ref{gbgg}), (\ref{apslp}) -- (\ref{apmap}):
\begin{eqnarray}
  \label{brbsnum}B(\bar B_s\to\gamma\gamma) &=& 1.23\cdot 10^{-6}\\
  \label{brbdnum}B(\bar B_d\to\gamma\gamma) &=& 3.11\cdot 10^{-8}
\end{eqnarray}
To display the size of power corrections from 1PI diagrams we compare
$A^p_+$ with $A^p_-$ in (\ref{apslp}) and (\ref{amslp}) for 
$B_s\to\gamma\gamma$
\begin{eqnarray}
&& A^u_+=4.87  \qquad\quad A^c_+=4.87 \label{apucnum} \\
&& A^u_-=5.29  \qquad\quad A^c_-=5.07 - 0.53\, i \label{amucnum}
\end{eqnarray}
The calculable power corrections, which manifest themselves in the difference
between $A^p_+$ and $A^p_-$, are of the expected canonical size of 
${\cal O}(10\%)$. The precise value of these power-suppressed
effects depends strongly on the renormalization scale $\mu$.
Typically, the relative size of the 1PI terms diminishes when the
scale is lowered, both due to an increase in $C_7$, as well as
due to the simultaneous decrease of the combination $C_1+ 3 C_2$.

The main errors in (\ref{brbsnum}--\ref{amucnum}) come from the variation 
of the nonperturbative input parameters $\lambda_B$ and the decay constants 
$f_{B_d}$ and $f_{B_s}$ which are all poorly known and enter the branching 
ratios quadratically. 
The residual scale dependence is sizeable as well because we calculated at 
leading logarithmic accuracy only. The variation of the branching ratio 
with the scale would be less severe if next-to-leading QCD corrections 
were known. We expect the NLL corrections to increase the branching ratio 
as was the case for both the inclusive \cite{CMM} and exclusive 
\cite{BB,BFS} $b\to s\gamma$ decays. 

The effect of next-to-leading logarithmic corrections for the branching
ratio of $b\to s\gamma$ can be reproduced by choosing a low
renormalization scale ($\mu\approx m_b/2$) in the leading logarithmic
expressions. If such a low scale were also relevant for 
$B\to\gamma\gamma$ the branching ratios would read
\begin{eqnarray}
  B(\bar B_s\to\gamma\gamma)_{\mu=m_b/2} &=& 1.53\cdot 10^{-6}\\
  B(\bar B_d\to\gamma\gamma)_{\mu=m_b/2} &=& 4.08\cdot 10^{-8}
\end{eqnarray}
However, we prefer to use the nominal $\mu=m_b$ and to quote the 
standard scale ambiguity with $m_b/2 <\mu < 2 m_b$ as a theoretical 
uncertainty.

For comparison we also show the results for the case where short-distance
QCD effects are neglected altogether. This amounts to taking $\mu=M_W$
in the leading-order Wilson coefficients and gives
\begin{eqnarray}
  B(\bar B_s\to\gamma\gamma)_{\mu=M_W} &=& 0.60\cdot 10^{-6}\\
  B(\bar B_d\to\gamma\gamma)_{\mu=M_W} &=& 1.22\cdot 10^{-8}
\end{eqnarray}
Clearly, the leading logarithmic QCD corrections yield a substantial
enhancement \citer{HI,CLY}
of the branching ratios (\ref{brbsnum}), (\ref{brbdnum}),
similar to the case of $b\to s\gamma$.

The numerical evaluation of (\ref{rCPpar}) gives for central values of the 
input parameters and at $\mu=m_b$ the following predictions for the 
CP-violating ratios $r^-_{CP}$:
\begin{eqnarray}
  \label{rcpbsnum}r^-_{CP}(B_s) &=&  0.35\%\\
  \label{rcpbdnum}r^-_{CP}(B_d) &=& -9.53\%
\end{eqnarray}
The effect is negligible for $B_s\to\gamma\gamma$ as expected.
For $B_d\to\gamma\gamma$ CP violation occurs at the level of
about $10\%$.
It is interesting to note that in $B\to\gamma\gamma$ a non-vanishing 
CP asymmetry appears already at ${\cal O}(\alpha_s^0)$. This is possible via 
a modified BSS mechanism: 
Instead of the usual QCD penguin, already the electroweak loop
in Fig.~\ref{fig:1PI} itself leads to a CP-conserving phase. 

\begin{figure}[t]
   \epsfxsize=12cm
   \centerline{\epsffile{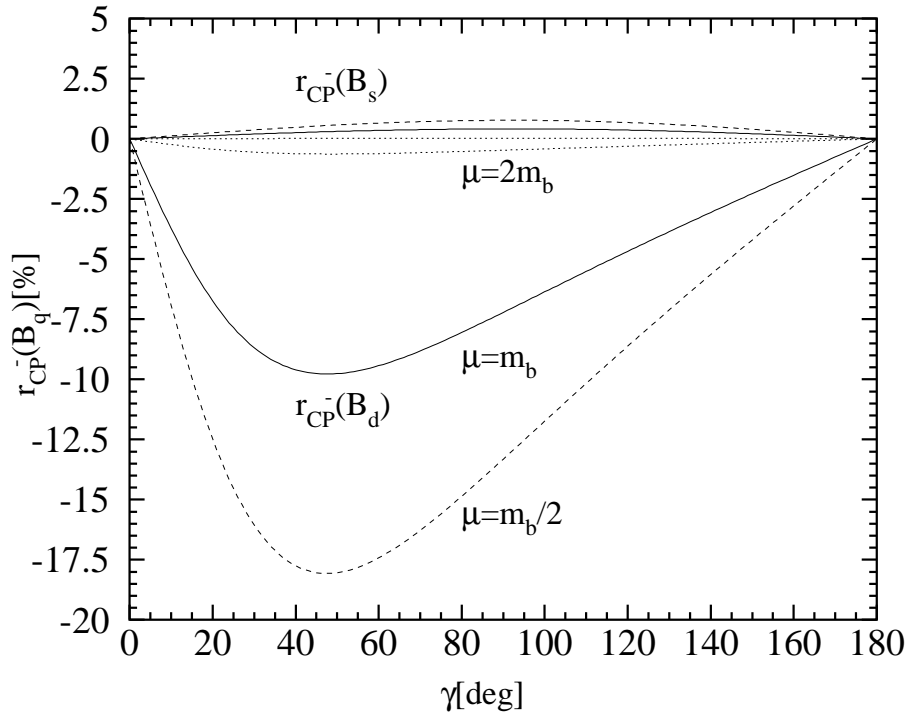}}
\caption{The CP-violating ratios $r^-_{CP}$ for decays of neutral $B_d$ 
  and $B_s$ mesons to two photons as a function of the CKM angle $\gamma$, 
  each for three values of the renormalization scale $\mu=m_b/2$, $m_b$ 
  and $2 m_b$.\label{fig:rCPgam}}
\end{figure}

The scale dependence of the CP asymmetry is rather strong,
as can be seen in Fig.  \ref{fig:rCPgam}.
The ratio $r_{CP}$ also depends sensitively on fundamental CKM parameters 
such as the CKM angle $\gamma$ (Fig.~\ref{fig:rCPgam}).
The extremal value of $r_{CP}^-(B_d)$ is obtained for 
$\gamma=50\,\mathrm{deg}$. 
The sensitivity of the branching ratios and CP asymmetries to variations 
in the relevant input parameters is summarized in Table~\ref{tab:output}.
\begin{table}[tpb]
\renewcommand{\arraystretch}{1.1}
\begin{center}
\begin{tabular}{|l|c|c|c|c|}
\hline\hline
   & $B(\bar B_d \to \gamma\gamma)$ & $B(\bar B_s\to \gamma\gamma)$ & 
  $r^-_{CP}(B_d)$ & $r^-_{CP}(B_s)$\\
   & $[10^{-8}]$ & $[10^{-6}]$ & $[\%]$ & $[\%]$\\
\hline\hline
central      & 3.11          & 1.23          & $-$9.53 & 0.35
\\
\hline\hline

$\lambda_B$      & +6.41/$-$1.58 & 2.45/$-$0.61  &  +4.04/$-$3.91
 & +0.14/$-$0.15\\
\hline

$f_B$        & +1.00/$-$0.86 & +0.34/$-$0.30 & -- & --\\
\hline

$\mu\in [m_b/2,2m_b]$ & +0.97/$-$0.66 & +0.31/$-$0.21 & +8.91/$-$8.07& 
              +0.30/$-$0.32\\
\hline

$m_c$       & +0.10/$-$0.12 & +0.03/$-$0.04 & +1.75/$-$1.51 & +0.05/$-$0.06\\
\hline

$\gamma=(58\pm 24)^\circ$ & +1.29/$-$0.96 & +0.01/$-$0.02  & +1.64/$-$0.25 & 
 +0.06/$-$0.12\\
\hline\hline
\end{tabular}
\end{center}
\caption[]{Predictions for branching ratios and CP asymmetries with the 
errors from the individual input uncertainties.\label{tab:output}}
\end{table}
As already mentioned, the dominant uncertainty for our prediction
of branching ratios comes from the variation of the hadronic
parameter $\lambda_B$, which enters the leading-power amplitude. 
So far, our value for $\lambda_B$ quoted in Table~\ref{tab:input}
is only an educated guess with the uncertainty taken appropriately
large. It would be highly desirable to get a better estimate
of this important parameter. We anticipate that this problem
can be addressed soon using QCD sum rules or lattice QCD.
As $\lambda_B$ parametrizes the first negative moment of the $B$ meson
wave function, it is a universal quantity that appears in
many exclusive decays. An experimental determination might also be 
possible using
radiative semileptonic $B$ decays, or even a future measurement of
$B\to\gamma\gamma$.
The variation of the remaining parameters, other than $\lambda_B$,
changes the branching ratios of $\bar B_d\to\gamma\gamma$ and
$\bar B_s\to\gamma\gamma$ by about $\pm 50\%$ and $\pm 35\%$,
respectively. If both branching fractions could be measured, one could
consider their ratio where the dominant uncertainties from
$\lambda_B$ and $f_B$ would cancel to a large extent. The main
uncertainty for the CP asymmetries at present comes from the
variation of the renormalization scale $\mu$. This could be reduced,
in principle, by including higher order QCD corrections.

Our predictions for the branching ratios are roughly two orders of magnitude 
below the current experimental bounds in (\ref{brbsdggex}). 
The recent upper limit for $B(B_d\to\gamma\gamma)$ from BaBar in 
(\ref{brbsdggex}) improved the previous limit from the 
L3 collaboration \cite{L3} by a factor of 20. 
With more integrated luminosity accumulated the upper bound will
be further reduced, but it will barely reach the interesting region.
The high-luminosity option SuperBaBar suggests a total integrated
luminosity of 10~ab$^{-1}$. The number of observed $B_d\to\gamma\gamma$
events is then expected to be about two dozens 
for a branching fraction of $3\cdot 10^{-8}$ \cite{SuperBaBar}.
Running such a machine at the $\Upsilon(5S)$ resonance would
open the possibility of measuring $B_s\to\gamma\gamma$ as well.
An interesting option might further be the operation
of a future $e^+e^-$ linear collider at the $Z$ resonance.

\section{Conclusions}\label{sec:concl}

In this paper we have presented a systematic discussion of the decays
$B_{s,d}\to\gamma\gamma$, exploiting the simplifications of QCD
in the heavy-quark limit. This enabled us to study these processes
for the first time within a model-independent framework.
In particular, we discussed the hierarchy of contributions to the decay
amplitude in powers of $\Lambda_{QCD}/m_b$. The leading contribution
to $B_s\to\gamma\gamma$ comes from the chromomagnetic $b\to s\gamma$
transition where the second photon is emitted from the $s$-quark
constituent.

We have also investigated the $B\to\gamma\gamma$ matrix elements
of four-quark operators $Q_1$ and $Q_2$, which proceed through 1PI
triangular up- and charm-quark loops. These are suppressed by
$\Lambda_{QCD}/m_b$ relative to the leading contribution. Disregarding
other terms of the same order, the consideration of these 1PI
diagrams is nevertheless meaningful because they are the leading effects
that induce a difference between the up- and charm-quark sectors of the
amplitude. This is important for CP violation. They are also
interesting as the partonic counterparts of certain, so-called
long-distance effects in $B\to\gamma\gamma$ 
or, similarly, $D\to\gamma\gamma$. We have argued that the 1PI matrix 
elements are calculable in QCD factorization and have explicitly
shown this property at leading (${\cal O}(\alpha^0_s)$) and
next-to-leading order (${\cal O}(\alpha_s)$) in the strong coupling.
Numerically the 1PI effect is only of order $10\%$ in the amplitude,
in accordance with a power-correction of canonical size.

Our estimates for branching ratios and CP asymmetries are given in 
Table~\ref{tab:output}. At present there are still very large
uncertainties in the absolute predictions for the branching fractions
due to the high sensitivity on the hadronic parameter $\lambda_B$,
which is a universal, process-independent property of the $B$ meson.
The quantity $\lambda_B$ is poorly studied at present, but an improved 
determination should be possible in the future. Also, a measurement of
$B_s\to\gamma\gamma$ could be a useful test of the hadronic dynamics
and for $\lambda_B$, which plays an important role for many other exclusive
$B$ decays.

The experimental study of $B_{s,d}\to\gamma\gamma$ will not be easy.
Presumably second generation $B$-factories or the high-intensity operation
of a future $e^+e^-$ linear collider at the $Z$ pole would offer the
best prospects. The radiative modes $B_{s,d}\to\gamma\gamma$ are interesting
probes of heavy-flavour physics in their own right and especially in the
context of factorization and the theory of exclusive $B$ decays.
In view of this, new ways towards their experimental observation
should continue to be pursued.

\vfill\eject

\end{document}